\documentstyle[12pt]{article}

\begin{document}

\def\nic{1}
\def\mnp{2}
\def\ati{3}
\def\dh{4}
\def\ipl{5}
\def\bkn{6}
\def\cbs{7}
\def\ind{8}
\def\fv{9}
\def\imb{10}

\begin{titlepage}

\begin{flushright}
{ CERN-TH 6471 \\
HU-TFT 92-15 \\}
\end{flushright}

\vskip 0.6truecm

\begin{center}
{\large \bf INDEX THEOREMS AND LOOP SPACE GEOMETRY }
\end{center}
  \par \vskip .1in \noindent

\begin{center}
{\bf A. Hietam\"aki} \\
\vskip 0.3cm
{\it Research Institute for Theoretical Physics,
University of Helsinki \\
Siltavuorenpenger 20 C, SF-00170 Helsinki, Finland}
\end{center}

\begin{center}
and
\end{center}

\begin{center}
{\bf Antti J. Niemi $^{*}$ } \\
\vskip 0.3cm
{\it Theory Division, CERN, CH-1211 Geneva 23, Switzerland $^{\dag}$ }
\end{center}

\vskip 1.5cm

We investigate the evaluation of the Dirac index using
symplectic geometry in the loop space of the
corresponding supersymmetric quantum mechanical model. In particular, we
find that if we impose a simple first class constraint, we can
evaluate the Callias index of an odd dimensional Dirac operator
directly from the quantum mechanical model which yields the Atiyah-Singer
index of an even dimensional Dirac operator in one more dimension. The
effective action obtained by BRST quantization of this
constrained system can be interpreted in terms of loop space symplectic
geometry, and the corresponding path integral for the index can
be evaluated exactly using the recently developed localization
techniques.
\vfill

\begin{flushleft}
\rule{5.1 in}{.007 in}\\
$^{*}$ {\small E-mail: ANIEMI@PHCU.HELSINKI.FI \\ }
$^{\dag}$ {\small permanent address: Research Institute for Theoretical
Physics, University of  Helsinki, Siltavuorenpenger 20 C, SF-00170
Helsinki, Finland } \\ \vskip 0.4cm
{ April 1992 \\
CERN-TH 6471 \\
HU-TFT 92-15}
\end{flushleft}

\end{titlepage}

\vfill\eject

{\it 1. Introduction.} There are many different ways to
characterize supersymmetric theories.  Maybe the best known is the
existence of a Nicolai transformation [\nic]. It identifies the fermion
determinant of a supersymmetric path integral as a Jacobian determinant for a
change of variables that takes the bosonic action into a Gaussian form.

Recently a conceptually different, geometric characterization
of supersymmetry has been proposed [\mnp]. This approach
originates from the ideas introduced in [\ati], [\dh],  and
is based on the formalism developed in [\ipl], [\bkn]. It
relates a supersymmetric theory to  symplectic geometry in
a superloop space, with half of the  bosonic and fermionic
variables viewed as superloop space  coordinates, and
the remaining half of these variables viewed as superloop space one-forms.
In this  superloop space the action of a
supersymmetric theory is locally exact with respect to a model
independent superloop space equivariant exterior derivative:
The noninteracting part of the action is a linear combination
of two terms, a loop space symplectic two-form and the contraction of the
corresponding symplectic one-form with a model independent
supervector field, that determines a natural action of
the  group SO(2). The interaction part of the action can
then be identified as a locally exact one-form in the
superloop space. This characterization of supersymmetry in terms of
(super)symplectic geometry appears to be at least as general as the
characterization in terms of a Nicolai transformation.  Indeed, in this
approach the Nicolai transformation  can be viewed as a loop space
change of variables to  (locally defined) Darboux variables.

In [\ati] the results of [\dh] are applied to
evaluate the Atiyah-Singer index of a
Dirac operator on a compact manifold: The supersymmetric quantum mechanical
path
integral which yields the index, localizes to the critical
points of the action corresponding to constant field configurations. The
correct
expression for the index then follows from the infinite dimensional
Duistermaat-Heckman integration formula [\dh]. In [\ipl] these observations are
combined with the localization technique developed in [\bkn], to evaluate
the Atiyah-Singer index of a Dirac operator on a compact manifold
and with arbitrary background gauge fields. When combined with the formalism
developed in [\mnp], these observations then suggest a potentially useful
method
for evaluating path integrals in generic supersymmetric theories.
In the present paper we shall be interested in developing such methods:
As a concrete example we shall consider
path integrals which are relevant for the
computation of the Callias index [\cbs]
of odd-dimensional Dirac operators. We explain how the localization
technique introduced in [\ipl,\bkn] generalizes to this case. Furthermore, in
addition we find that the  Callias index of an odd dimensional Dirac operator
can also be obtained from the quantum
mechanical path integral that yields the Atiyah-Singer index of a related Dirac
operator in one more dimensions. This connection between the index formulas
in even and odd dimensions emerges, when we consider the even dimensional
quantum mechanical model together with a constraint that
effectively eliminates one space dimension. The BRST quantization of the
ensuing constrained system yields an effective action, which admits an
interpretation in terms of loop space symplectic geometry. The evaluation
of the corresponding path integral using the
localization technique introduced in
[\ipl,\bkn] then yields the correct expression
for the index in odd dimensions.

\vskip 0.6cm

{\it 2. D=1 Index Theorem:} We shall first illuminate our
general approach by investigating the evaluation of the index of
the one-dimensional Dirac hamiltonian
$$
{\cal H} ~=~ i \sigma^{1} {d \over dq} + \sigma^{2} W_{q}
{}~=~ \left(\matrix{ 0 & D
\cr D^{\dag} & 0 \cr}\right)
\eqno (1)
$$
Here
$$
D ~=~ i {d \over dq} - i W_{q} (q)
\eqno (2)
$$
and the derivative of the superpotential $W(q)$ has a soliton
profile,
$$
W_{q}(\pm\infty) ~=~ a_{\pm}
\eqno (3)
$$
The corresponding supersymmetric quantum mechanical model is
obtained by identifying the operators $D$ and  $D^{\dag}$ as
generators of a canonical supersymmetry algebra,
$$
\left(\matrix{ 0 & D \cr 0 & 0 \cr}\right)
{}~\rightarrow~ Q ~=~ \frac{1}{\sqrt{2}} \eta (p + i W_{q})
\eqno (4.a)
$$
$$
\left(\matrix{ 0 & 0 \cr D^{\dag} & 0 \cr}\right) ~\rightarrow~
Q^{\dag} ~=~ \frac{1}{\sqrt{2}} \bar\eta (p - i W_{q})
\eqno (4.b)
$$
where $\eta$ and $\bar\eta$ are anticommuting variables, and
the nontrivial Poisson brackets  are
$$
\{ p , q \} ~=~ \{ \bar\eta , \eta \} ~=~ 1
\eqno (5)
$$
The supersymmetry algebra is
$$
\{ Q , Q^{\dag} \} ~=~ H
\eqno (6.a)
$$
$$
\{ Q , Q \} ~=~ \{ Q^{\dag} , Q^{\dag} \} ~=~ \{ Q , H \} ~=~
\{ Q^{\dag} , H \} ~=~ 0
\eqno (6.b)
$$
with
$$
H ~=~ \{ Q , Q^{\dag} \} ~=~ \frac{1}{2} p^{2} + \frac{1}{2}
W_{q}^{2} + i \bar\eta W_{qq} \eta
\eqno (7)
$$
The index of the operator (1) is obtained in the $T\to
\infty$ limit of the superpartition function
$$
Z ~=~ Tr \{ e^{iT H} \} ~=~ \int [dpdq] [d\bar\eta d\eta] exp\{ i
S_{B} + iS_{F} \}
\eqno (8)
$$
with
$$
S_{B} + S_{F} ~=~ \int\limits_{0}^{T} p \dot q + {\bar\eta} \dot
\eta -  \{ Q , Q^{\dag} \}
\eqno (9)
$$

In order interpret (8), (9) in terms of superloop space symplectic geometry, we
introduce the canonical transformation
$$
p ~\to~ e^{-\Phi} p e^{\Phi} ~=~ p + \{ p , \Phi\} + \frac{1}{2\!}
\{ \{ p , \Phi\}, \Phi \} + ~ ...
\eqno (10.a)
$$
$$
q ~\to~ e^{-\Phi} q e^{\Phi} ~=~ q + \{ q , \Phi\} + \frac{1}{2\!}
\{ \{  q , \Phi\}, \Phi \} + ~ ...
\eqno (10.b)
$$
If we select
$$
\Phi(q) ~=~ - i \int\limits^{q} W(q') dq'
\eqno (11)
$$
we get
$$
p ~ {\buildrel {\Phi} \over {\longrightarrow} } ~ p - iW(q)
\eqno (12.a)
$$
$$
q ~ {\buildrel {\Phi} \over {\longrightarrow} } ~ q \eqno
(12.b)
$$
For the generators of the supersymmetry algebra this yields
$$
Q ~ { \buildrel {\Phi} \over {\longrightarrow} } ~
e^{-\Phi} Q e^{\Phi}
{}~=~ \frac{1}{\sqrt{2}} \eta p
\eqno (13.a)
$$
$$
Q^{\dag} ~ { \buildrel {\Phi} \over {\longrightarrow} } ~ e^{-\Phi} Q^{\dag}
e^{\Phi} ~=~ \frac{1}{\sqrt{2}} \bar\eta (p - 2i W(q))
\eqno (13.b)
$$
and for the action
$$
S_{B} + S_{F} ~\to~ \int\limits_{0}^{T} p \dot q +
\bar\eta\dot\eta  - \frac{1}{2}p^{2} + i p W_{q}(q) + i
\bar\eta W_{qq}(q) \eta
\eqno (14)
$$
Defining
$$
\eta ~\to~ \frac{1}{\sqrt{2}}(\theta_{1} + i \theta_{2})
\eqno (15.a)
$$
$$
\bar\eta ~\to~ \frac{1}{\sqrt{2}}(\theta_{1} - i \theta_{2})
\eqno (15.b)
$$
$$
p ~\to~ -ip + \dot q
\eqno (15.c)
$$
we then get
$$
S_{B} + S_{F} ~=~ \int\limits_{0}^{T} \frac{1}{2} {\dot q}^{2} +
\frac{1}{2} p^{2} - p W_{q} + \frac{1}{2} (\theta_{1}\dot
\theta_{1} + \theta_{2} \dot\theta_{2}) - \theta_{2}
W_{qq}\theta_{1}
\eqno (16)
$$

In this form the action can be directly interpreted in terms of
superloop space symplectic geometry. For this we consider a superloop space
with $q(t)$ and $\theta_{2}(t)$ viewed as superloop space coordinates,  and
$\theta_{1}(t)$ and $p(t)$  as one-forms.  The exterior derivative is
$$
{\bf d} ~=~ \theta_{1} {\delta \over
\delta q } + p {\delta \over \delta \theta_{2}}
\eqno (17)
$$
and if we introduce the superloop space symplectic one-form
$$
\vartheta ~=~ - \frac{1}{2} {\dot q} \theta_{1} + \frac{1}{2} \theta_{2}
p
\eqno (18)
$$
and the corresponding symplectic two-form which
determines superloop space Poisson
brackets is the exterior derivative of (18),
$$
\Omega ~=~ {\bf d} \vartheta ~=~  \frac{1}{2} p^{2} + \frac{1}{2}
\theta_{1} \dot \theta_{1}
\eqno (19)
$$
and if we define the vector field
$$
{\bf i}_{S} ~=~ - {\dot q} \cdot {\bf i}_{\theta_{1} } - {\dot \theta_{2}}
\cdot
{\bf i}_{p}
\eqno (20)
$$
we get the noninteracting part of the supersymmetric
quantum mechanics action (16) as
$$
S ~=~ ({\bf d} + {\bf i}_{S}) \vartheta ~=~ \Omega + {\bf
i}_{S}\vartheta ~=~ \int\limits_{0}^{T} \frac{1}{2} {\dot
q}^{2} + \frac{1}{2} p^{2}
+ \frac{1}{2}\theta_{1} \dot \theta_{1} + \frac{1}{2}\theta_{2} \dot
\theta_{2}
\eqno (21)
$$
The interaction is obtained by
defining the superloop space scalar {\it i.e.} zero-form
$$
{\cal W}  ~=~ \theta_{2} W_{q}
\eqno (22)
$$
Since the interior multiplication of a loop space vector field and
a loop space scalar vanishes,
$$
{\bf i}_{S} {\cal W} ~=~ 0
\eqno (23)
$$
we then find that in the superloop space the action (16) can
be represented as ($p\to -p$)
$$
S_{B} + S_{F} ~=~ ({\bf d}+{\bf i}_{S})(\vartheta + {\cal W}) \\
=~ \int\limits_{0}^{T} \frac{1}{2} {\dot q}^{2} + \frac{1}{2}
p^{2} + p W_{q} + \frac{1}{2}(\theta_{1} \dot \theta_{1} + \theta_{2} \dot
\theta_{2}) + \theta_{1} W_{qq} \theta_{2}
\eqno (24)
$$
Notice in particular, that this action is a linear combination
of exact forms with degree zero, one and two.

We shall now proceed to the evaluation of the path integral (8), (24) using
the localization techniques developed in [\ipl,\bkn].  For this we
first observe that for the superloop space Lie derivative,
$$
{\cal L}_{S} ~=~ {\bf d} {\bf i}_{S} + {\bf i}_{S} {\bf d} ~=~ -
{\dot q} \partial_{q} - {\dot \theta}_{2} \partial_{\theta_{2}} -
{\dot p} {\bf i}_{p} - {\dot \theta}_{1} {\bf i}_{\theta_{1}} ~=~ -
\partial_{t}
\eqno (25)
$$
Obviously
$$
{\cal L}_{S} ({\dot q}\theta_{1}) ~=~ 0
\eqno (26)
$$
Hence we conclude from [\ipl,\bkn] that the path integral
(8), (24) remains intact if we
redefine\footnotemark\footnotetext{ Notice that as
explained in [\ipl], we can not redefine the second
term in (18) in the same manner.
Due to the constant mode, such a redefinition is not a small, local variation.}
$$
{\vartheta} ~\to~ \vartheta_{\beta} ~=~ - \frac{\beta}{2}
{\dot q}\theta_{1} + \frac{1}{2} \theta_{2}p
\eqno (27)
$$
For the action this yields
$$
S_{B} + S_{F} ~\to~ ({\bf d}+{\bf i}_{S})(\vartheta_{\beta} +
{\cal W}) \\ =~ \int\limits_{0}^{T} \frac{\beta}{2} {\dot
q}^{2} + \frac{1}{2} p^{2} + p W_{q} +
\frac{\beta}{2}\theta_{1} \dot \theta_{1} + \frac{1}{2}\theta_{2} \dot
\theta_{2} + \theta_{1} W_{qq} \theta_{2}
\eqno (28)
$$
and from the arguments presented in [\bkn] we conclude that the path
integral with (28) is $\beta$-independent.

We define the path integral measure in (8) as [\ipl]
$$
[dq][d\theta_{1}][dp][d\theta_{2}] ~=~ dq_{0}d\theta_{10}dp_{0}
d\theta_{20} \prod_{t}dq_{t}d\theta_{1t}dp_{t}d\theta_{2t}
\eqno (29)
$$
Here we have defined
$$
q(t) ~=~ q_{0} ~+~ q_{t}
{}~~~~~~~~~~~
\theta_{1}(t) ~=~ \theta_{10} ~+~ \theta_{1t}
\eqno (30.a)
$$
$$
p(t) ~=~ p_{0} ~+~ p_{t}
{}~~~~~~~~~~~
\theta_{2}(t) ~=~ \theta_{20} ~+~ \theta_{2t}
\eqno (30.b)
$$
with $q_{0}$, $\theta_{10}$, $p_{0}$, $\theta_{20}$ the
constant modes. We then introduce the change of
variables
$$
q(t) ~\rightarrow~ q_{0} ~+~ \frac{1}{\sqrt{\beta}} q_{t}
\eqno (31.a)
$$
$$
\theta_{1}(t) ~\rightarrow~ \theta_{10} ~+~
\frac{1}{\sqrt{\beta}} \theta_{1t}
\eqno (31.b)
$$
The pertinent Jacobian in the path integral measure (29) is
trivial. Since the path integral is independent of $\beta$
we can take the $\beta\to\infty$ limit which yields for the
action in (28)
$$
S_{B} + S_{F} ~\to~ \int\limits_{0}^{T}\frac{1}{2} {\dot q}_{t}^{2}
+ \frac{1}{2}p_{0}^{2} + \frac{1}{2}p_{t}^{2} + \frac{1}{2}
\theta_{1t}{\dot \theta}_{1t} + \frac{1}{2}\theta_{2t}{\dot
\theta}_{2t} + \theta_{10}W_{qq}(q_{0})\theta_{20} + p_{0}
W_{q}(q_{0}) + {\cal O}(\frac{1}{\sqrt{\beta}})
\eqno (32)
$$
Integrating over $\theta_{1}$, $\theta_{2}$ and $p$ we
then get for the path integral
$$
Z ~=~ \sqrt{ \frac{T}{2\pi} }
\int\limits_{-\infty}^{\infty} dq_{0} d\theta_{10} d\theta_{20} exp \{
- \frac{i}{2} T W_{q}^{2}(q_{0}) + i T \theta_{10} W_{qq}(q_{0}) \theta_{20} \}
\eqno (33.a)
$$
$$
= ~ \frac{1}{2\pi}\int\limits_{-\infty}^{\infty}
dq_{0} W_{qq}(q_{0}) exp \{ - \frac{i}{2} T W_{q}^{2}(q_{0}) \}
\eqno (33.b)
$$
and in the $T\to\infty$ limit this yields for the index of the
operator (1)
$$
Z ~{\buildrel{T\to\infty}\over{\longrightarrow}}~ Index ({\cal H}) ~=~
DimKer (D) - DimKer (D^{\dag}) ~=~  \frac{1}{2}\left[
\frac{a_{+}}{|a_{+}|} - \frac{a_{-}}{|a_{-}|} \right]
\eqno (34)
$$

This result agrees with the exact evaluation
in [\cbs,\ind], and verifies that in the present case the
localization technique developed in [\ipl,\bkn] is applicable.

\vskip 0.6cm

{\it 3. Relation to the Atiyah-Singer Index.} In [\ind] it has been shown
how the index of an odd dimensional Dirac operator can be related to the index
of an operator defined in one more dimensions by an
infinite space version of the
Atiyah-Patodi-Singer index theorem. Here, we shall obtain a somewhat different
connection between even and odd space index theorems: We find
that the index of a Dirac operator can be
evaluated from a supersymmetric quantum mechanical model
which yields the index of a related Dirac operator in one more dimension.
This result is a consequence of a very simple constrained quantization
procedure: We select the higher dimensional operator in such a way that it
coincides with the lower dimensional operator when we demand translation
invariance in one space direction. At the level of quantum mechanical path
integrals we impose this translation invariance as a first class
constraint, and as a consequence we find that the
constrained path integral yields
the index  of the lower dimensional operator. In particular, in this way we can
derive the Callias index theorem directly from the Atiyah-Singer
index theorem of a higher dimensional Dirac operator.

As an example, we shall evaluate the one dimensional index formula (34) using
a supersymmetric path integral which yields the two
dimensional Atiyah-Singer index. For this, we consider the following
two dimensional Dirac operator
$$
{\cal D} ~=~ i
\gamma^{i}\partial_{i} - \gamma^{i}A_{i}
\eqno (35)
$$
Here we use the Pauli-matrix representation $\gamma^{1}
=\sigma_{y}$ and $\gamma^{2} = \sigma_{x}$ of the
$\gamma$-matrices. Notice that
if we set $A_{1} = 0$ and demand that $A_{2}$ depends only
on the variable $q_{1} \sim q$, the $q_{2}$-derivative becomes
quite irrelevant. By ignoring this derivative and identifying
$A_{2} \sim W_{q}$
the operator (35) then essentially reduces to the operator (1).
We shall now demonstrate, that if we
implement these conditions as a canonical first class constraint in the
pertinent supersymmetric quantum mechanical model, the path integral evaluation
of the Atiyah-Singer index for (35) yields the Callias index (34) of the one
dimensional operator (1).

The canonical
supersymmetry generator corresponding to (35) is
$$
Q ~=~ \frac{1}{\sqrt{2}}\psi^{i}(p_{i} - A_{i})
\eqno (36)
$$
and the Poisson bracket supersymmetry algebra is
$$
\{ Q , Q \} ~=~ H ~=~ \frac{1}{2} (p_{i} - A_{i})^{2} + \frac{1}{2}
\psi^{i} F_{ij}\psi^{j}
\eqno (37)
$$

In [\ipl] it has been shown that the index of (35) can be obtained by
applying the localization technique to the canonical path integral
with action
$$
S ~=~ \int\limits_{0}^{T} p_{i}\dot q_{i} +
\frac{1}{2}\psi_{i} {\dot \psi}_{i} - \frac{1}{2}(p_{i} -
A_{i})^{2} - \frac{1}{2}\psi^{i}F_{ij}\psi^{j}
\eqno (38)
$$
In order for this to yield the index of the operator (1), we introduce in
(38) the abelian first class constraint
$$
p_{2} ~\sim~ 0
\eqno (39)
$$
If the gauge fields $A_{i}$ are functions of $q_{1}$
only, this constraint commutes with the supersymmetry generator (36),
$$
\{ p_{2} , Q \} ~=~0
\eqno (40)
$$
Consequently it also commutes with the hamiltonian (37) and we have a first
class constrained system.

We shall enforce the constraint (39) in the path integral
using the BRST gauge fixing procedure. For this, following [\fv]
we introduce a pair of canonically conjugated Grassman variables
with Poisson brackets
$$
\{ {\cal P} , \eta \} ~=~ \{ {\bar {\cal P}} , {\bar \eta} \} ~=~
-1
\eqno (41)
$$
and commuting Lagrange multiplier variables with
$$
\{ \pi , \lambda \} ~=~ -1
\eqno (42)
$$
The BRST operator corresponding to (39) is then
$$
\Omega ~=~ \eta p_{2} + \bar\eta \pi
\eqno (43)
$$
In order to fix the gauge in the path integral, we introduce the gauge fermion
$$
\Psi ~=~ \frac{1}{\beta} {\bar {\cal P}} q_{2} + {\cal P}\lambda
\eqno (44)
$$
where $\beta$ is a parameter. This yields the BRST
gauge-fixed action [\fv]
$$
S~=~ \int\limits_{0}^{T} p_{i} \dot q_{i} + \frac{1}{2} \psi_{i}
{\dot \psi}_{i} + \pi {\dot \lambda} + {\cal P} {\dot \eta} + {\bar
{\cal P}} {\dot {\bar \eta}} - \frac{1}{2} (p_{i} - A_{i})^{2}
- \frac{1}{2}\psi^{i}F_{ij}\psi^{j} - \frac{1}{\beta} \eta{\bar
{\cal P}} - {\bar \eta}{\cal P} - \lambda p_{2} - \frac{1}{\beta}
\pi q_{2}
\eqno (45)
$$
We change the variables $\pi \to \beta \pi$ and ${\bar {\cal
P}} \to \beta {\bar{\cal P}}$. The corresponding Jacobian is
trivial, and since the path integral is $\beta$ independent [\fv] we
can take the $\beta \to \infty$ limit. In this limit we obtain for
the action (45)
$$
S ~\rightarrow~ \int\limits_{0}^{T} p_{i}\dot q_{i} + \frac{1}{2}
\psi_{i} {\dot \psi}_{i} - \frac{1}{2} (p_{i} - A_{i})^{2} -
\frac{1}{2} \psi_{i} F_{ij}\psi_{j} + \lambda p_{2} + \pi q_{2}
\eqno (46)
$$
Eliminating $p_{i}$, $\pi$ we get further
$$
S ~\rightarrow~ \int\limits_{0}^{T} \frac{1}{2} {\dot q}_{1}^{2} +
\frac{1}{2} \psi_{i} {\dot \psi}_{i} + \frac{1}{2} \lambda^{2} +
{\dot q}_{1} A_{1} + \lambda A_{2} - \frac{1}{2}
\psi_{i}F_{ij}\psi_{j}
\eqno (47)
$$
and the pertinent path integral is subject to the condition $q_{2} = 0$,
which is enforced by a $\delta$-functional.

The gauge fixed action (47) admits an interpretation in terms of
loop space equivariant cohomology in a superloop space with
coordinates $q_{1}$ and $\psi_{2}$, and with $\psi_{1}$ and
$\lambda$ the corresponding superloop space one-forms. For this
we define the superloop space  exterior derivative
$$
{\bf d} ~=~ \psi_{1} \partial_{q_{1}} ~+~ \lambda
\partial_{\psi_{2}}
\eqno (48)
$$
and vector field
$$
{\bf i}_{S} ~=~ - {\dot q}_{1} {\bf i}_{\psi_{1}} - {\dot
\psi}_{2} {\bf i}_{\lambda}
\eqno (49)
$$
The corresponding Lie-derivative is
$$
{\cal L}_{S} ~=~ {\bf d} {\bf i}_{S} + {\bf i}_{S} {\bf d} ~=~ -
\partial_{t}
\eqno (50)
$$
Defining the one-form
$$
\vartheta ~=~ - \frac{1}{2} {\dot q}_{1}\psi_{1} + \frac{1}{2}
\psi_{2} \lambda - A_{1} \psi_{1}
\eqno (51)
$$
and the zero-form
$$
{\cal W} ~=~ A_{2}\psi_{2}
\eqno (52)
$$
we then find that the action (47) can be represented as
$$
S ~=~ ({\bf d} + {\bf i}_{S}) (\vartheta + {\cal W})
\eqno (53)
$$

We shall now proceed to evaluate the corresponding path integral using the
localization technique developed in [\ipl,\bkn]. For this we first re-define
the
one-form (51) into
$$
\vartheta ~\to~  - \frac{\beta}{2} {\dot q}_{1}\psi_{1} +
\frac{1}{2} \psi_{2} \lambda - A_{1} \psi_{1}
\eqno (54)
$$
According to [\ipl,\bkn] the path integral is independent of $\beta$.
Proceeding as in [\ipl] and (28)-(33) we then find in the
$\beta\to\infty$ limit for the action
$$
iS ~\to~ i\int\limits_{0}^{T} \frac{1}{2} {\dot q}_{1t}^{2} +
\frac{1}{2} {\psi_{it}} {\dot \psi}_{it} + \frac{1}{2}
\lambda_{0}^{2} + \frac{1}{2}\lambda_{t}^{2} + \lambda_{0}
A_{2}(q_{10}) - \frac{1}{2} \psi_{i0}F_{ij}(q_{10})\psi_{j0} ~+~
{\cal O}(\frac{1}{\sqrt{\beta}})
\eqno (55.a)
$$
$$
\rightarrow ~ - \frac{i}{2}T A_{2}^{2}(q_{10}) -
\frac{i}{2}T \psi_{i0}F_{ij}(q_{10})\psi_{j0}
\eqno (55.b)
$$
where we have again decomposed the variables in
their constant and non-constant modes, used the
$\delta$-function in the functional integral to
eliminate the variable $q_{2}$, and
in the final step eliminated the nonconstant modes.

The evaluation of the remaining  $\psi_{i0}$ and $q_{10}$ integrals
proceeds as in the case of the two-dimensional Atiyah-Singer index
for (35) [\ipl]. Indeed, by identifying $A_{2} \sim W_{q}$ and $A_{1} \sim 0$
we
conclude that the actions in (33.a) and (55.b) coincide, hence we find again
the
result (34). Consequently we have found a connection between the two
dimensional
Atiyah-Singer index of (35) and the one
dimensional Callias index of (1): The path
integral evaluation of the former yields the latter if we impose translation
invariance in the $q_{2}$ direction as an abelian first-class constraint in the
supersymmetric model (38).

\vskip 0.6cm

{\it 4. Index Theorem in Higher Dimensions: } We shall now proceed to
generalize the previous results for Dirac operators
defined in general D-dimensional (flat) spaces.
For this, we first consider the supersymmetry generators
discussed in [\imb],
$$
Q ~=~ \frac{1}{2} (\theta_{i} p_{i} + {\bar \theta}_{i} W_{i})
\eqno (56.a)
$$
$$
{\bar Q} ~=~ \frac{1}{2} (\bar\theta_{i} p_{i} - \theta_{i} W_{i})
\eqno (56.b)
$$
where the Poisson brackets of the anticommuting variables are
$$
\{ \theta_{i} , \theta_{j} \} ~=~ \{ {\bar\theta}_{i} ,
{\bar\theta}_{j} \} ~=~ \delta_{ij}
\eqno (57)
$$
and the supersymmetry algebra is
$$
\{ Q , Q \} ~=~ \{ {\bar Q} , {\bar Q} \} ~=~ H ~=~
\frac{1}{2} p_{i}^{2} + \frac{1}{2} W_{i}^{2} + \bar\theta_{i}
\theta_{j} W_{ij}
\eqno (58.a)
$$
$$
\{ Q , {\bar Q} \} ~=~  \{ Q , H \} ~=~ \{ {\bar Q} , H \} ~=~ 0
\eqno (58.b)
$$
As explained in [\imb], this supersymmetry algebra is relevant to a
class of Dirac operators whose index has been considered in [\cbs].
The corresponding canonical supersymmetry action is
$$
S ~=~ \int\limits_{0}^{T} p_{i} \dot q_{i} + \frac{1}{2}
\theta_{i} {\dot \theta}_{i} + \frac{1}{2} {\bar\theta}_{i} {\dot
{\bar\theta}}_{i} - \frac{1}{2} p_{i}^{2} - \frac{1}{2} W_{i}^{2}
- {\bar \theta}_{i}\theta_{k} W_{ik}
\eqno (59)
$$
In order to interpret (59) in terms of symplectic geometry in a superloop space
we again introduce a canonical conjugation such that
$$
p_{i} ~\rightarrow~ p_{i} - i W_{i}
\eqno (60.a)
$$
$$
q_{i} ~\rightarrow~ q_{i}
\eqno (60.b)
$$
Generalizing (10)-(16) we then find for the action
$$
S ~\to~ \int\limits_{0}^{T} \frac{1}{2} {\dot q}_{i}^{2} +
\frac{1}{2} \theta_{i} {\dot \theta}_{i} + \frac{1}{2} {\bar
\theta}_{i} {\dot {\bar\theta}}_{i} + \frac{1}{2}p_{i}^{2} -
p_{i} W_{i} - \bar\theta_{i} W_{ik}\theta_{k}
\eqno (61)
$$
which generalizes (16) to D dimensions. The
corresponding superloop space interpretation is obtained by
introducing the equivariant exterior derivative
$$
{\bf d} + {\bf i}_{S} ~=~ \theta_{i} \partial_{q_{i}} - p_{i}
\partial_{{\bar\theta}_{i}} - {\dot q}_{i} {\bf i}_{\theta_{i}}
+ {\dot {\bar\theta}}_{i} {\bf i}_{p_{i}}
\eqno (62)
$$
With the one-form
$$
\vartheta ~=~ - \frac{1}{2} {\dot q}_{i} \theta_{i} - \frac{1}{2}
{\bar\theta}_{i} p_{i}
\eqno (63)
$$
and zero-form
$$
{\cal W} ~=~ \bar\theta_{i} W_{i}
\eqno (64)
$$
we can then write the action (61) as
$$
S ~=~ ({\bf d}+ {\bf i}_{S}) (\vartheta + {\cal W})
\eqno (65)
$$

The evaluation of the path integral proceeds as
before: We introduce a generalization of (27) to localize the path
integral to constant configurations, and repeating the steps that
led to (33) we then find that the path integral reduces
to  finite dimensional integral over the bosonic zero mode
$$
Z ~=~  \int \prod\limits_{i} dq_{i0} d{\bar\theta}_{i0} d\theta_{i0}
exp \{ - \frac{i}{2} T W_{i}^{2}(q_{i0}) - i T {\bar\theta}_{i0} W_{ik}(q_{i0})
\theta_{k0} \}
\eqno (66.a)
$$
$$
=~ \sqrt{(\frac{T}{2\pi})^{n}}\int \prod\limits_{i} dq_{i0}
det || W_{ij}(q_{i0})|| exp \{ - \frac{i}{2} T W_{i}^{2}(q_{i0}) \}
\eqno (66.b)
$$
Notice that this is a direct generalization of (33) to D
dimensions. The result equals that obtained in [\imb], and yields in the
$T\to\infty$ limit the index of the corresponding  Dirac operator.

\vskip 0.5cm

{\it 5. Relation to the Atiyah-Singer Index.} In analogy with
(35)-(55) we can relate (66) to the Atiyah-Singer index of a
higher dimensional Dirac operator. For this, we need to select the higher
dimensional operator in such a manner that it reduces to the one corresponding
to (56) if we impose translation invariance in some of the coordinate
directions. The ensuing constrained path integral quantization of the higher
dimensional supersymmetry action then yields the index the original
Dirac operator, obtained in the $T\to\infty$ limit of (66). As an example we
shall here apply the path integral evaluation of the four dimensional
Atiyah-Singer index theorem described in [\ipl] to compute the index  of a
three
dimensional Dirac hamiltonian coupled to a SU(2) Yang-Mills-Higgs system,
$$
{\cal H} ~=~ \alpha^{k}(i
\partial_{k} + A_{k}) - \beta \Phi
\eqno (67)
$$
where $A_{k} = A_{k}^{\alpha}T^{\alpha}$
is a hermitean gauge field in an isospin-J representation of SU(2),
$$
[T^{\alpha} , T^{\beta} ] ~=~ i \epsilon^{\alpha\beta\gamma}
T^{\gamma}
{}~~~~~~\&~~~~ Tr \{ T^{\alpha} T^{\beta} \} ~=~
J(J+1)\delta^{\alpha\beta}
\eqno (68)
$$
and we choose $T^{3}$ to be diagonal with eigenvalues
$J,J-1,...,-J$.

For a monopole background the gauge field $A_{k}$ can be viewed as a compact
perturbation. As a consequence in the index computation it is usually customary
to consider only the Higgs field background
[\cbs]. With $\Phi^{a} \sim W_{i}$ the
ensuing canonical supersymmetry algebra is
then of the form (56), and the index can
be evaluated as explained above.  Here we shall
proceed to describe, how this index
can also be obtained from the path integral evaluation of the four dimensional
Atiyah-Singer index for the Dirac operator
$$
D ~=~ \gamma^{\mu} (i \partial_{\mu} - A_{\mu})
\eqno (69)
$$

If we identify $A_{0}\sim \Phi$ in (67) and require
translation invariance in the $\mu=0$ direction, (69)
essentially reduces to (67).
Consequently, we expect that if we impose translation invariance as a first
class constraint in the canonical supersymmetry algebra corresponding to (69),
the path integral evaluation of the Atiyah-Singer index for (69) subject to
this
constraint yields the index of (67). For this, we introduce the four
dimensional
canonical supersymmetry algebra corresponding to (69) with
$$
Q ~=~ \frac{1}{\sqrt{2}}\psi_{\mu} \left( p_{\mu} - A_{\mu}^{\alpha}
T^{\alpha}(\varphi) \right)
\eqno (70)
$$
Here we have realized the gauge generators $T^{\alpha}$
canonically on a co-adjoint
orbit: With $\varphi^{a}$ coordinates on the co-adjoint orbit and
$\omega_{ab}(\varphi)$ the symplectic two-form corresponding to
the given representation of the gauge group, we represent the Lie
algebra canonically, in terms of the co-adjoint orbit Poisson brackets
$$
\{ T^{\alpha}(\varphi) , T^{\beta}(\varphi) \} ~=~ \partial_{a}
T^{\alpha} \omega^{ab} \partial_{b} T^{\beta} ~=~
\epsilon^{\alpha\beta\gamma}T^{\gamma}(\varphi)
\eqno (71)
$$

In order for (70) to represent the operator (67) we demand that
it is subject to the first class constraint
$$
p_{0} ~\sim~0
\eqno (72)
$$
The condition
$$
\{ p_{0} , Q \} ~=~0
\eqno (73)
$$
then implies that $A_{\mu}$ is independent of the coordinate
$q_{0}$, and if we identify $A_{0} \sim \Phi$ we obtain the
desired constrained canonical realization of the operator (67).
The corresponding supersymmetry algebra is determined by
$$
\{ Q , Q \} ~=~ H ~=~ \frac{1}{2} (p_{\mu} -
A_{\mu}^{\alpha}T^{\alpha})^{2} + \frac{1}{2}
\psi_{\mu} F_{\mu\nu}^{\alpha}T^{\alpha} \psi_{\nu}
\eqno (74)
$$
and the action which will yield the index of (67) is
$$
S~=~ \int\limits_{0}^{T} p_{\mu} {\dot q}_{\mu} +
\theta_{a} {\dot \varphi}^{a} +
\frac{1}{2} \psi_{\mu} {\dot \psi}_{\mu} - \frac{1}{2} (p_{\mu} -
A_{\mu}^{\alpha}T^{\alpha})^{2} - \frac{1}{2} \psi_{\mu}
F_{\mu\nu}^{\alpha}T^{\alpha}\psi_{\nu} - \frac{1}{2}
c^{a}\omega_{ab}c^{b}
\eqno (75)
$$
and this action is subject to the constraint (72).
Notice that except for the constraint, this action coincides with that used
in [\ipl] to evaluate the Atiyah-Singer index of (69). Here
$c^{a}$ are anticommuting variables introduced to exponentiate the square-root
determinant of the symplectic structure on the co-adjoint orbit, which appears
as a measure factor in the canonical path integral [\ipl].

We again enforce the constraint (72) using BRST
gauge-fixing. Introducing the corresponding gauge fermion
(44) and generalizing the steps (41)-(47) we then find for the action
$$
S ~\to~ \int\limits_{0}^{T} \{ \frac{1}{2} \dot q_{k}^{2}
+ \frac{1}{2} \psi_{\mu}{\dot \psi}_{\mu} + \frac{1}{2} \lambda^{2} +
\theta_{a}\dot\varphi^{a} + {\dot q}_{k} A_{k}  + \lambda \Phi
- \frac{1}{2} \psi_{\mu} F_{\mu\nu}^{\alpha}T^{\alpha}
\psi_{\nu} - \frac{1}{2} c^{a}\omega_{ab} c^{b} \}
\eqno (76)
$$
Here $\lambda$ is the Lagrange multiplier field for the
constraint (72), and the path integral is subject to the
$\delta$-function constraint $q_{0} = 0$.

In order to interpret (76) in the superloop space we define
$$
{\bf d} ~=~ \psi_{k}
\partial_{q_{k}} + \lambda \partial_{\psi_{0}} + c^{a}
\partial_{\varphi^{a}}
\eqno (77)
$$
$$
{\bf i}_{S} ~=~ - {\dot q}_{k} {\bf i}_{\psi_{k}} - {\dot
\psi}_{0} {\bf i}_{\lambda} - {\dot \varphi}^{a} {\bf i}_{c^{a}}
\eqno (78)
$$
Notice that again we find for the Lie derivative
$$
{\cal L}_{S} ~=~ {\bf d} {\bf i}_{S} + {\bf i}_{S} {\bf d} ~=~
-\partial_{t}
\eqno (79)
$$
Defining the one-form
$$
\vartheta ~=~ - \frac{1}{2} {\dot q}_{k} \psi_{k} + \frac{1}{2}
\psi_{0} \lambda - \theta_{a}c^{a} -
A_{k}^{\alpha}T^{\alpha} \psi_{k}
\eqno (80)
$$
and the zero-form
$$
{\cal W} ~=~ \psi_{0} A_{0} ~\sim~ \psi_{0} \Phi
\eqno (81)
$$
we then find for the action (76) the representation
$$
S ~=~ ({\bf d}+ {\bf i}_{S}) (\vartheta + {\cal W})
\eqno (82)
$$

We evaluate the path integral using the localization
technique developed in [\ipl,\bkn]. For this we introduce
the following one-parameter
generalization of (80),
$$
\vartheta ~\to~ - \frac{\beta}{2} {\dot
q}_{k} \psi_{k} + \frac{\beta}{2} \psi_{0} \lambda -
\theta_{a}c^{a} - A_{k}^{\alpha}T^{\alpha} \psi_{k}
\eqno (83)
$$
The general arguments presented in [\ipl,\bkn] again imply that the path
integral is independent of the parameter $\beta$.
We again divide the variables into their constant and non-constant
modes, and using the $\beta$ independence of the path integral we
then find in the $\beta\to\infty$ limit for the action
$$
iS ~\to~ i\int\limits_{0}^{T} \{ \frac{1}{2} \dot q_{kt}^{2} + \frac{1}{2}
\psi_{\mu t}\dot\psi_{\mu t} + \frac{1}{2}
\lambda_{0}^{2} +
\theta_{a}\dot\varphi^{a} + \lambda_{0} \Phi(q_{0})
- \frac{1}{2}\psi_{\mu 0} F_{\mu\nu}^{\alpha}(q_{0}) T^{\alpha}
\psi_{\nu 0} - \frac{1}{2} c^{a} \omega_{ab} c^{b} \}
\eqno (84)
$$
and by integrating over the nonconstant modes we finally get
$$
iS ~\to~  \frac{i}{2} T \Phi^{\alpha}\Phi^{\beta}
T^{\alpha}T^{\beta} - \frac{i}{2} T \psi_{0\mu}
F_{\mu\nu}\psi_{0\nu}
\eqno (85)
$$
Notice that this is a direct generalization of our
earlier results (33.a), (55.b)
and (66.a). Notice also that this action differs from the corresponding action
obtained in the evaluation of the Atiyah-Singer index theorem [\ipl]: The
second
term in (85) does coincide with that relevant for the computation of the four
dimensional Atiyah-Singer index theorem, but the first
term in (85) is a consequence
of the constraint (72). In the $T\to\infty$ limit only the second term  in (85)
survives, and we get the index of the operator (68) [\cbs,\ind]
$$
Z~{\buildrel {T\to\infty} \over {\longrightarrow} } ~
\frac{1}{8\pi} J(J+1) \oint dS^{i} B^{\alpha}_{i} \frac{\Phi^{\alpha}}{|\Phi|}
\eqno (86)
$$

\vskip 0.5cm

{\it 6. In conclusions}, we have investigated the localization
technique developed in [\ipl,\bkn] for the evaluation of supersymmetric path
integrals by computing the index of certain Dirac operators. In each case, the
localization technique appears to yield the correct result for the path
integral. In particular, we have found that the method also yields a novel
point of view to index theorems for odd-dimensional Dirac operators: The index
can be computed directly from the path integral relevant to a higher
dimensional Atiyah-Singer index, by introducing a simple first class
constraint that eliminates the extra dimensions. The ensuing BRST quantized
canonical action admits a superloop space interpretation and as a consequence
the localization techniques developed in [\ipl,\bkn] become directly
applicable.

\vskip 1.0cm

\noindent
We thank K. Palo for discussions.

\vfill\eject

{\bf References}

\vskip 0.8cm

\begin{enumerate}

\item H. Nicolai, Phys. Lett. {\bf 89B} (1980) 341; Nucl. Phys. {\bf B176}
(1980) 419

\item A. Yu. Morozov, A.J. Niemi and K. Palo, Physics Letters {\bf B271}
(1991) 365; Nucl. Phys. {\bf B} (to appear)

\item M.F. Atiyah, Asterisque {\bf 131} (1985) 43; J.-M. Bismut, Comm. Math.
Phys. {\bf 98} (1985) 213; {\it ibid.} {\bf 103} (1986) 127

\item J.J. Duistermaat and G.J. Heckman, Inv. Math. {\bf 72} (1983) 153;
M.F. Atiyah and R. Bott, Topology {\bf 23} (1984) 1

\item A. Hietam\"aki, A.Yu. Morozov, A.J. Niemi and K. Palo, Phys. Lett. {\bf
B263} (1991) 417

\item M. Blau, E. Keski-Vakkuri and A.J. Niemi, Phys. Lett. {\bf B246} (1990)
92; A.J. Niemi and P. Pasanen, Phys. Lett. {\bf B253} (1991) 349; E.
Keski-Vakkuri, A.J. Niemi, G. Semenoff and O. Tirkkonen,
Phys. Rev. {\bf D44} (1991) 3899

\item C. Callias, Comm. Math. Phys. {\bf 62} (1978) 213; R.
Bott and R. Seeley, Comm. Math. Phys. {\bf 62} (1978) 235

\item A.J Niemi and G.W. Semenoff, Nucl. Phys. {\bf B269} (1986) 131

\item E.S. Fradkin and G.A. Vilkovisky, Phys. Lett. {\bf B55} (1975) 224; A.J.
Niemi, Phys. Repts. {\bf 184} (1989) 147

\item C. Imbimbo and S. Mukhi, Nucl. Phys. {\bf B242} (1984)
81

\end{enumerate}

\end{document}